\documentclass[twocolumn,british,english,prb,superscriptaddress]{revtex4-1}
\usepackage[T1]{fontenc}
\usepackage[latin9]{inputenc}
\usepackage{babel}
\usepackage{units}
\usepackage{textcomp}
\usepackage{amsmath}
\usepackage{amssymb}
\usepackage{graphicx}
\usepackage[unicode=true,pdfusetitle,
 bookmarks=true,bookmarksnumbered=false,bookmarksopen=false,
 breaklinks=false,pdfborder={0 0 1},backref=false,colorlinks=false]
 {hyperref}
\usepackage{breakurl}

\makeatletter

\providecommand{\tabularnewline}{\\}

 \@ifundefined{textcolor}{}
 {%
  \definecolor{BLACK}{gray}{0}
  \definecolor{WHITE}{gray}{1}
  \definecolor{RED}{rgb}{1,0,0}
  \definecolor{GREEN}{rgb}{0,1,0}
  \definecolor{BLUE}{rgb}{0,0,1}
  \definecolor{CYAN}{cmyk}{1,0,0,0}
  \definecolor{MAGENTA}{cmyk}{0,1,0,0}
  \definecolor{YELLOW}{cmyk}{0,0,1,0}
  }

\RequirePackage{colortbl, tabularx}
\@ifundefined{comment}{}
  {
   
  }%

\makeatother

\usepackage{babel}

\makeatother

\begin{document}

\title{Fluence dependent femtosecond quasi-particle and Eu$^{2+}$-spin
relaxation dynamics in EuFe$_{2}$(As,P)$_{2}$}

\author{A. Pogrebna}

\affiliation{Complex Matter Dept., Jozef Stefan Institute, Jamova 39, Ljubljana,
SI-1000, Ljubljana, Slovenia }

\author{T. Mertelj}

\email{tomaz.mertelj@ijs.si}

\selectlanguage{english}%

\affiliation{Complex Matter Dept., Jozef Stefan Institute, Jamova 39, Ljubljana,
SI-1000, Ljubljana, Slovenia }

\affiliation{CENN Nanocenter, Jamova 39, Ljubljana SI-1000, Slovenia}

\author{G. Cao}

\affiliation{Department of Physics, Zhejiang University, Hangzhou 310027, People\textquoteright s
Republic of China}

\author{Z. A. Xu}

\affiliation{Department of Physics, Zhejiang University, Hangzhou 310027, People\textquoteright s
Republic of China}

\author{D. Mihailovic}

\affiliation{Complex Matter Dept., Jozef Stefan Institute, Jamova 39, Ljubljana,
SI-1000, Ljubljana, Slovenia }

\affiliation{CENN Nanocenter, Jamova 39, Ljubljana SI-1000, Slovenia}

\date{\today}

\begin{abstract}
We investigated temperature and fluence dependent dynamics of the
time resolved optical reflectivity in undoped spin-density-wave (SDW)
and doped superconducting (SC) EuFe$_{2}$(As,P)$_{2}$ with emphasis
on the ordered Eu$^{2+}$-spin temperature region. The data indicate
that the SDW order coexists at low temperature with the SC and Eu$^{2+}$-ferromagnetic
order. Increasing the excitation fluence leads to a thermal suppression
of the Eu$^{2+}$-spin order due to the crystal-lattice heating while
the SDW order is suppressed nonthermally at a higher fluence.
\end{abstract}
\maketitle

\section{Introduction}

In the iron-based superconductors family\cite{KamiharaKamihara2006,kamiharaWatanabe2008}
EuFe$_{2}$(As,P)$_{2}$\cite{RenTao2009} and Eu(Fe,Co)$_{2}$As$_{2}$\cite{JiangXing2009}
offer an interesting experimental possibility to study the competition
between the ferromagnetic (FM) and superconducting (SC) order parameters
that can lead to nonuniform magnetic and SC states\cite{AndersonSuhl1959,BuzdinBulaevskii1984,JiangXing2009,BlachowskiRuebenbauer2012},
since the optimal critical temperature $T\mathrm{_{c}\sim28}$ K\cite{JeevanKasinathan2011}
is comparable to the Eu$^{2+}$-spin ordering temperatures $T\mathrm{_{M}\sim20}$
K.\cite{RenZhu2008,RenTao2009} 

While no coherent picture of Eu$^{2+}$-spin ordering upon P or Co
doping exists\cite{JiangXing2009,ZapfWu2011,ZapfJeevan2013,NandiJin2014},
a pure FM ordering\emph{\cite{NandiJin2014}} coexisting with superconductivity
was reported by Nandi \emph{et al.\cite{NandiJin2014}} in EuFe$_{2}$(As$_{0.85}$P$_{0.15}$)$_{2}$\emph{.}
Our recent transient magneto-optical spectroscopy study\cite{PogrebnaMertelj2015}
also points towards the simple FM Eu$^{2+}$-spin order in the superconducting
EuFe$_{2}$(As$_{1-x}$P$_{x}$)$_{2}$ with a slow energy transfer
between the FeAs-plane quasiparticles and Eu$^{2+}$ spins indicating
a weak magnetic-dipole dominated coupling between the SC and FM order
parameters. 

Here we extend our previous transient reflectivity study\cite{PogrebnaMertelj2015}
first focusing briefly on the spin-density-wave dominated part of
the phase diagram followed by a study of the superconducting phase-diagram
region at varying excitation density to study a suppression of the
coexistent orders on an ultrafast timescale.

\section{Experimental}

\subsection{Samples}

Single crystals of EuFe$_{2}$(As$_{1-x}$P$_{x}$)$_{2}$ were grown
by the flux method, similar to the previous report\cite{JiaoTao2011}.
Small Eu chunks and powders of Fe, As and P (Alfa Aesar, > 99.9\%)
were mixed together in the molar ratio of Eu:Fe:As:P = 1:5:5(1-$x$):5$x$
($x$ is the nominal P content) and sealed in an evacuated quartz
ampule. After heating the mixture up to 973 K for 24 hours, the obtained
precursor was thoroughly ground before being loaded into an alumina
crucible. The crucible was then sealed by arc wielding in a tube made
of stainless steel under atmosphere of argon, and then heated up to
1573 K over 10 hours in a muffle furnace filled with argon. After
holding at 1573 K for 5 hours, the furnace was cooled down to 1223
K at the rate of 5 K/h. followed by switching off the furnace. Large
crystals with size up to 4\texttimes 4\texttimes 0.6 mm$^{3}$ could
be harvested. 

The as-grown crystals were characterized by X-ray diffraction, which
showed good crystallinity as well as single \textquotedblleft 122\textquotedblright{}
phase. The exact composition of the crystals was determined by energy
dispersive X-ray spectroscopy affiliated to a field-emission scanning
electron microscope (FEI Model SIRION). The measurement precision
was better than 5\% for the elements measured.

\begin{figure}[tbh]
\begin{centering}
\includegraphics[angle=-90,width=0.95\columnwidth]{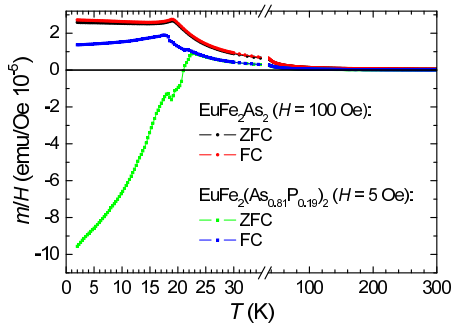} 
\par\end{centering}

\caption{(Color online) Magnetic moment along the $c$-axis as a function of
temperature for both, field cooled (FC) and zero field cooled (ZFC)
cases.}

\label{fig:figSusc} 
\end{figure}

The out-of-plane magnetic susceptibilities shown in Fig. \ref{fig:figSusc}
are consistent with previous results. \cite{JiangLuo2009,ZapfJeevan2013}
From the susceptibility we infer Eu$^{2+}$ spin ordering temperatures
$T\mathrm{_{N}}=19$ K and $T_{\mathrm{C}}=17.6$ K in EuFe$_{2}$As$_{2}$
(Eu-122) and EuFe$_{2}$(As$_{0.81}$P$_{0.19}$)$_{2}$ (EuP-122),
respectively. EuP-122 also shows the onset of superconductivity at
$T\mathrm{_{c}=}22.7$ K.

\subsection{Optical setup}

Measurements of the photoinduced reflectivity, $\Delta R/R$, were
performed using the standard pump-probe technique, with 50 fs optical
pulses from a 250-kHz Ti:Al$_{2}$O$_{3}$ regenerative amplifier
seeded with an Ti:Al$_{2}$O$_{3}$ oscillator. We used the pump photons
with both, the laser fundamental ($\hbar\omega_{\mathrm{P}}=1.55$
eV) and the doubled ($\hbar\omega_{\mathrm{P}}=3.1$ eV) photon energy,
and the probe photons with the laser fundamental $\hbar\omega_{\mathrm{pr}}=1.55$
eV photon energy. When using the doubled photon energy the scattered
pump photons were rejected by long-pass filtering, while an \foreignlanguage{british}{analyzer}
oriented perpendicularly to the pump beam polarization was used for
rejection in the case of the degenerate pump and probe photon energies.
The pump and probe beams were nearly perpendicular to the cleaved
sample surface (001) with polarizations perpendicular to each other
and oriented with respect to the crystals to obtain the maximum/minimum
amplitude of the sub-picosecond $\Delta R/R$ at low temperatures.
The pump beam diameters were, depending on experimental conditions,
in 50-100 $\mu$m range with somewhat smaller probe beam diameters. 

\section{Results}

\subsection{Anisotropy of the $\Delta R/R$ transients}

\begin{figure}[tbh]
\begin{centering}
\includegraphics[angle=-90,width=0.98\columnwidth]{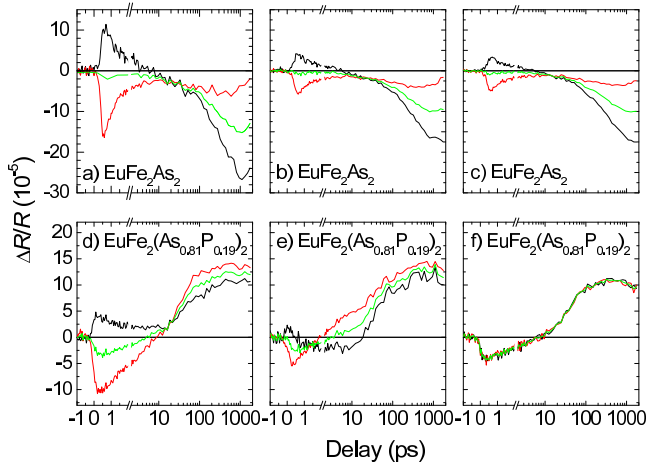} 
\par\end{centering}

\caption{(Color online) Variation of the transients across the sample surface
at $T=1.5$ K and 1.5-eV pump-photon energy in EuFe$_{2}$As$_{2}$
(a), (b), (c) and EuFe$_{2}$(As$_{0.81}$P$_{0.19}$)$_{2}$ (d),
(e), (f). Black and red lines are the transients for $\mathcal{P}^{+}$
and $\mathcal{P}^{-}$ probe polarizations, respectively, while green
lines are the corresponding averages. The pump fluences used were
$\mathcal{F}\sim10$ $\mu$J/cm$^{2}$(a) and $\mathcal{F}\sim3$
$\mu$J/cm$^{2}$ (b)-(f).}

\label{fig:figAniso}
\end{figure}

At low temperatures we observe a 2-fold rotational anisotropy of the
response with respect to the probe polarization at both doping levels.
In the absence of information about the in-plane crystal axes orientation
we denote the probe-polarization orientation according to the polarity
of the observed sub-picosecond low-$T$ response as $\mathcal{P}^{+}$
and $\mathcal{P}^{-}$. 

In EuP-122 we found a significant variation of the anisotropy, as
well as the transients shape, across the sample surface. On the other
hand, as shown in Fig. \ref{fig:figAniso}, there is almost no variation
of the $\mathcal{P}^{+}$, $\mathcal{P}^{-}$ averaged transients
indicating that the anisotropy variation is due to the twin domain
structure on the length scale of the probe-beam diameter of $\sim$50
$\mu$m. For all other measurements we therefore measured the single
domain response by choosing the position on the sample surface with
maximal anisotropy.

\subsection{Response in the SDW state}

\begin{figure}[tbh]
\begin{centering}
\includegraphics[angle=-90,width=0.98\columnwidth]{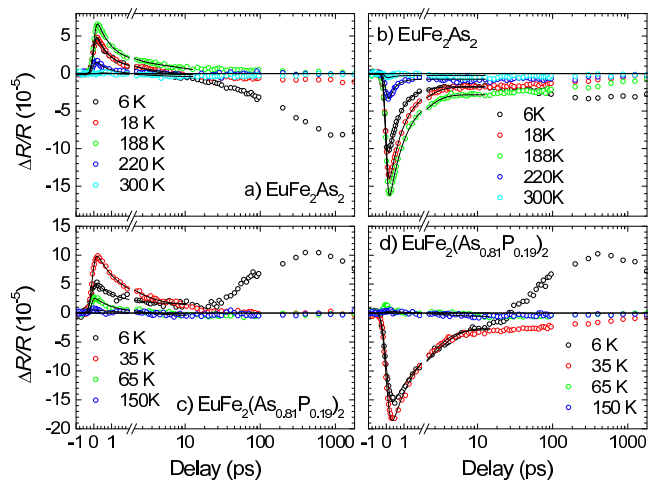} 
\par\end{centering}

\caption{(Color online) Photoinduced reflectivity transients at representative
temperatures at $\mathcal{F}\sim10$ $\mu$J/cm$^{2}$ and 3.1 eV
pump photon energy in EuFe$_{2}$As$_{2}$ (a), (b) and EuFe$_{2}$(As$_{0.81}$P$_{0.19}$)$_{2}$
(c), (d). Left and right panels correspond to $\mathcal{P}^{+}$ and
$\mathcal{P}^{-}$ polarizations, respectively. The thin lines are
the finite-excitation-pulsewidth double-exponential relaxation fits.\cite{StojchevskaMertelj2012} }

\label{fig:figDrVsT} 
\end{figure}

\subsubsection{Experimental data}

\begin{figure}[tbh]
\begin{centering}
\includegraphics[angle=-90,width=0.98\columnwidth]{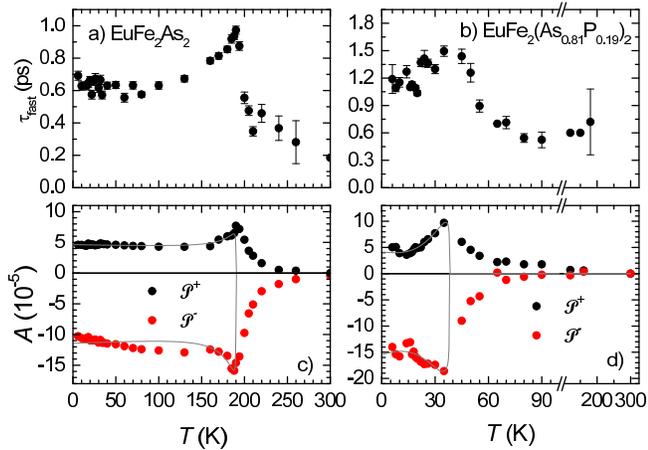} 
\par\end{centering}

\caption{(Color online) The relaxation time (a), (b) and amplitude (c), (d)
of the sub-picosecond response as a function of temperature in EuFe$_{2}$As$_{2}$
, (a), (c) and EuFe$_{2}$(As$_{0.81}$P$_{0.19}$)$_{2}$, (b), (d)
at $\mathcal{F}\sim10$ $\mu$J/cm$^{2}$ and 3.1 eV pump photon energy.
The full lines in are the bottleneck model\cite{StojchevskaMertelj2012,PogrebnaVujicic2014}
fits discussed in text.}

\label{fig:figAvsTfast} 
\end{figure}

In Fig. \ref{fig:figDrVsT} we show $\Delta R/R$ transients at a
few characteristic temperatures for both samples. Starting at $T=300$
K we observe sub-picosecond isotropic transients in both samples consistent
with previous results in related iron-based pnictides.\cite{StojchevskaMertelj2012,StojchevskaKusar2010,MerteljKusar2010}
With decreasing $T$ the 2-fold anisotropy appears below $\sim250$
K in Eu-122 and $\sim190$ K in EuP-122. The appearance of the anisotropy
is followed by a strong increase of the amplitude of the sub-picosecond
response peaking near the onset of the Fe-$d$ SDW order at 188 K
in Eu-122 and at significantly lower $T\sim35$ K in EuP-122 as shown
in Fig. \ref{fig:figAvsTfast}. The inital-picosecond-relaxation decay
time, obtained from double-exponential fits shown in Fig. \ref{fig:figDrVsT},
shows a divergence-like peak at the magneto-structural SDW transition
in Eu-122 while in EuP-122 it only shows a plateau with no peak, concurrent
with the amplitude maximum around $\sim35$ K. In Eu-122 the initial
picosecond relaxation also appears almost twice faster than in EuP-122.

\subsubsection{Discussion}

The initial fast relaxation in the undoped SDW state has been \foreignlanguage{british}{analysed}
previously in terms of the magnon-bottleneck model.\cite{StojchevskaKusar2010,StojchevskaMertelj2012,PogrebnaVujicic2014}
The data in Eu-122 are consistent with our broad-band probe results
on the same samples,\cite{PogrebnaVujicic2014} with the bottleneck-fit
parameters shown in Table \ref{tbl:gaps}.

\begin{table}
\medskip{}

\begin{tabular}{c|cc}
sample & $\nicefrac{2\Delta(0)}{k_{\mathrm{B}}T\mathrm{_{SDW}}}$ & $g_{\mathrm{ph}}$\tabularnewline
\hline 
EuFe$_{2}$As$_{2}$ & $12\pm7$ & $2.2\pm1$\tabularnewline
EuFe$_{2}$(As$_{0.81}$P$_{0.19}$)$_{2}$ & $5\pm1$ & $2.9\pm1.7$\tabularnewline
\end{tabular}

\caption{The SDW charge gap magnitudes and the relative effective number of
involved bosons as obtained from the fits to the data, described in
detail in Ref. [\onlinecite{PogrebnaVujicic2014}{]} and shown in
Fig. \ref{fig:figAvsTfast} (c), (d).}
\label{tbl:gaps}
\end{table}

The $T$\foreignlanguage{british}{-dependent} relaxation in doped
EuP-122 is qualitatively similar to Eu-122 suggesting the presence
of the magnetostructural/SDW transition at $T_{\mathrm{SDW}}\sim$35
K, consistent with the reported phase diagrams.\cite{JeevanKasinathan2011,TokiwaHubner2012,NandiJin2014}
The decrease of the relative charge gap magnitude $\nicefrac{2\Delta(0)}{k_{\mathrm{B}}T\mathrm{_{SDW}}}$
with SDW suppression upon P doping is similar as in the Ba(Fe,Co)$_{2}$As$_{2}$.\cite{StojchevskaMertelj2012}
Despite the effective SDW induced charge-gap magnitude in EuP-122
($2\Delta(0)\approx15$ meV, see Table \ref{tbl:gaps}) falls well
into the phonon energy range no significant increase of the relative
number of bottleneck bosons is observed suggesting that the electron-phonon
coupling in the vicinty of the SDW induced gap is weak.

In both samples we observe similarly to Ba(Fe,Co)$_{2}$As$_{2}$\cite{StojchevskaMertelj2012}
a 4-fold symmetry breaking at temperatures significantly exceeding
the corresponding magnetostructural transition temperatures which
we associate with nematic fluctuations\cite{ChuAnalytis2010,ChuAnalytis2010prb,DuszaLucarelli2011}.

\subsection{Response related to Eu$^{2+}$ spin ordering}

Concurrently with the Eu$^{2+}$-spin ordering\cite{xiaoSu2009,ZapfWu2011}
below $\sim17-19$ K we observe\cite{PogrebnaMertelj2015} in both
samples appearance of another much slower relaxation component with
a risetime of $\sim1$ ns in Eu-122 and $\sim100$ ps in EuP-122 (at
$T=1.5$ K) and the decay time beyond the experimental delay range
(see Fig. \ref{fig:figDrVsT}). In Eu-122 the slow component is rather
anisotropic, while in EuP-122 it appears almost isotropic at the chosen
pump fluence. 

Since the $T$- and $B$-dependence of this component have been already
analysed and discussed previously\cite{PogrebnaMertelj2015} we omit
further details here focusing on further aspects of our data not discussed
elsewhere.

\subsubsection{Pump fluence dependence}

\begin{figure}[tbh]
\begin{centering}
\includegraphics[angle=-90,width=0.95\columnwidth]{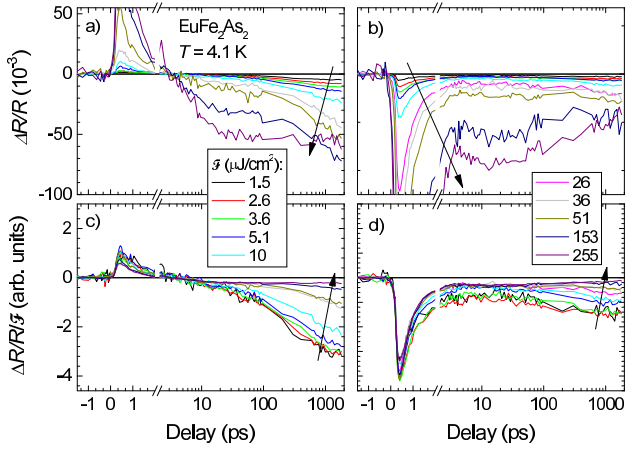} 
\par\end{centering}

\caption{(Color online) The low-$T$ transients in EuFe$_{2}$As$_{2}$ at
different fluences for $\mathcal{P}^{+}$, (a), and $\mathcal{P}^{-}$,
(b) probe polarizations and 1.55-eV pump photon energy. To emphasize
the low-$\mathcal{F}$ region the $\mathcal{F}$-normalized transients
from panels (a) and (b) are shown in (c) and (d), respectively. The
arrows indicate the direction of increasing fluence. Note that the
overlap of the $\mathcal{F}$-normalized scans indicates a linear
dependence.}

\label{fig:figDRvsFEFA1eV55} 
\end{figure}

\begin{figure}[tbh]
\begin{centering}
\includegraphics[angle=-90,width=0.98\columnwidth]{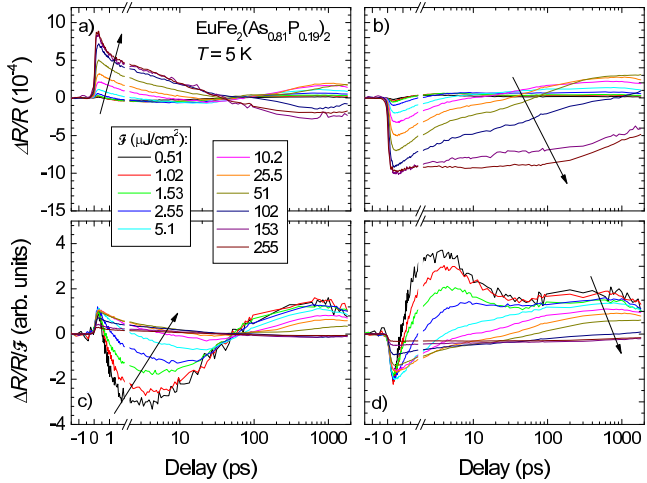} 
\par\end{centering}

\caption{(Color online) The low-$T$ transients in EuFe$_{2}$(As$_{0.81}$P$_{0.19}$)$_{2}$
at different fluences for $\mathcal{P}^{+}$, (a), and $\mathcal{P}^{-}$,
(b), probe polarizations and 1.55-eV pump photon energy. To emphasize
the low-$\mathcal{F}$ region the $\mathcal{F}$-normalized transients
from panels (a) and (b) are shown in (c) and (d), respectively. Arrows
indicate the direction of increasing fluence. Note that the overlap
of the $\mathcal{F}$-normalized scans indicates a linear dependence.}

\label{fig:figDRvsFEFAP1eV55} 
\end{figure}

\begin{figure}[tbh]
\begin{centering}
\includegraphics[angle=-90,width=0.9\columnwidth]{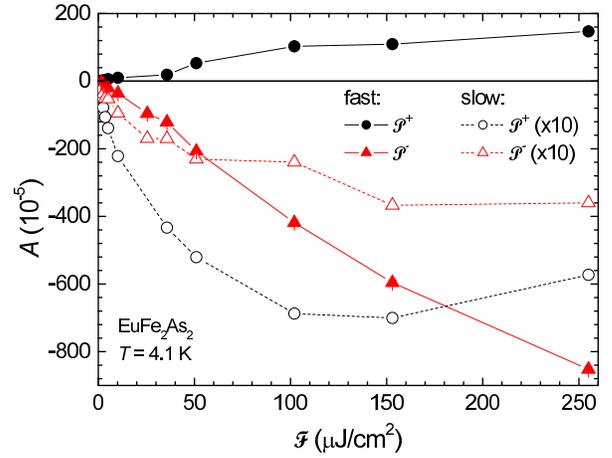} 
\par\end{centering}

\caption{(Color online) The amplitude of the short delay extrema and the amplitude
at the longest delay of the transients in EuFe$_{2}$As$_{2}$ as
functions of fluence at $T=6$ K and 1.55-eV pump photon energy.}

\label{fig:figAvsFEFA1eV55}
\end{figure}

The pump fluence dependence of the response at low $T$ is shown in
Figs. \ref{fig:figDRvsFEFA1eV55} and \ref{fig:figDRvsFEFAP1eV55}
for Eu-122 and EuP-122, respectively. While in Eu-122 the sub-picosecond
response shows only a slightly sublinear $\mathcal{F}$-dependence
(see Fig. \ref{fig:figAvsFEFA1eV55}) with almost $\mathcal{F}$-independent
sub-picosecond relaxation time in the full fluence range, the slow
part of the response shows a clear sublinear $\mathcal{F}$-dependence
already at $\mathcal{F}=\sim10$ $\mu$J/cm$^{2}$. The saturation
above $\mathcal{F}=\sim100$ $\mu$J/cm$^{2}$ is concurrent with
an increase of the risetime beyond the experimental delay range of
1.7 ns at the highest fluences. Interestingly, independently of $\mathcal{F}$
all $\mathcal{P}^{+}$ transients cross zero at $\sim3$-ps delay.
Above $\mathcal{F}\sim5$0 $\mu$J/cm$^{2}$ an additional dynamics
with a risetime on $\sim10$-ps and decay on a few-100-ps timescale
becomes apparent.

\begin{figure}[tbh]
\begin{centering}
\includegraphics[angle=-90,width=0.9\columnwidth]{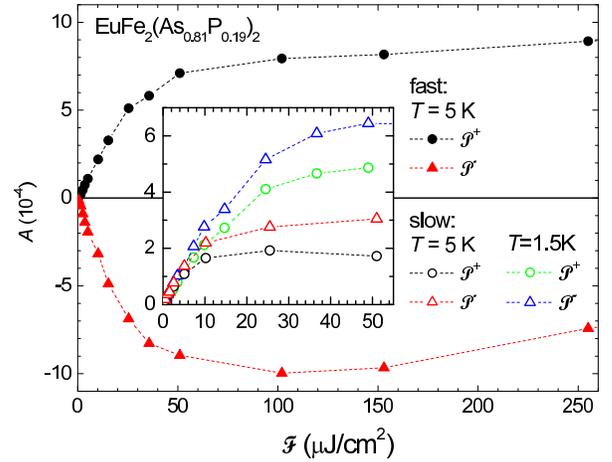} 
\par\end{centering}

\caption{(Color online) The amplitude of the sub-picosecond extrema of the
transients in EuFe$_{2}$(As$_{0.81}$P$_{0.19}$)$_{2}$ as functions
of fluence at $T=5$ K and 1.55-eV pump photon energy. The inset shows
the amplitude of the long delay extrema at low fluences including
the data measured at 1.5 K. }

\label{fig:figAvsFEFAP1eV55}
\end{figure}

In superconducting EuP-122 the $\mathcal{F}$-dependence of the transients
appears even more complex. There is a marked nonlinear behavior in
both, the picosecond and nanosecond responses. The amplitude of the
initial fast response with a sub-ps risetime is linear up to $\sim20$
$\mu$J/cm$^{2}$ and clearly saturates above $\sim50$ $\mu$J/cm$^{2}$
(see Fig. \ref{fig:figAvsFEFAP1eV55}). 

The shape of the fast response, contrary to Eu-122, strongly depends
on the fluence. For the $\mathcal{P}^{-}$ polarization at the lowest
$\mathcal{F}$ we observe after the initial sub-ps negative transient
an increase of the signal with zero crossing to an intermediate value
on a timescale of 2 ps followed by a partial decay on a $\sim10$-ps
timescale with no further decay within our time window (see Fig. \ref{fig:figDRvsFEFAP1eV55}
). With increasing $\mathcal{F}$ the initial increase becomes slower
with zero crossing moving beyond $\sim100$ ps, while the 10-ps partial
decay vanishes above $\mathcal{F}\sim5$$\mu$J/cm$^{2}$. 

For the $\mathcal{P}^{+}$ polarization the initial-few-ps response
at low $\mathcal{F}$ is similar to the $\mathcal{P}^{-}$ polarization
response with the opposite sign. At longer delays, beyond $\sim10$
ps, however, the slow component is observed causing a second zero
crossing with a characteristic risetime of $\sim200$ ps at the lowest
$\mathcal{F}$, which increases beyond the experimental delay range
with increasing $\mathcal{F}$. As a result, $\Delta R/R$ appears
rather isotropic at long delays. 

The slow response is linear in $\mathcal{F}$ up to $\sim10$ $\mu$J/cm$^{2}$
at $T=1.5$ K showing saturation with increasing $\mathcal{F}$ (see
inset to Fig. \ref{fig:figAvsFEFAP1eV55}). The amplitude of the initial
fast relaxation saturates above $\mathcal{F\sim}50$ $\mu$J/cm$^{2}$.
Concurrently it slows down (Fig. \ref{fig:figDRvsFEFAP1eV55}) and
starts to overlap with the onset of the slow one. 

Above $\mathcal{F\sim}100$ $\mu$J/cm$^{2}$ the shape of the slow
response shows an apparent qualitative change without the long delay
zero crossings in both polarizations resulting in a change of sign
of the nanosecond-timescale transient reflectivity.

\subsubsection{Discussion}

On the basis of a detailed temperature and magnetic field dependence
we have shown previously \cite{PogrebnaMertelj2015} that the slow
component can be associated with Eu$^{2+}$ spin demagnetization in
both the AFM and FM states. The saturation of the slow response at
high excitation fluences can therefore be associated with a complete
suppression of the Eu$^{2+}$ sublattice magnetizations due to the
lattice temperature rise above the magnetic ordering temperatures.
Taking into account the heat capacity\cite{JeevanHossain2008} and
optical\cite{WuBarisic2009} data we estimate\footnote{We obtain the light penetration depth of $\lambda_{\mathrm{opt}}=27$
nm at $\hbar\omega=1.55$ eV .} that the surface transient temperature rise due to the photoexcitation
is $\sim20$ K at $\mbox{\ensuremath{\mathcal{F}}}=10$ $\mu$J/cm$^{2}$.
This is consistent with the observed fluence dependence in EuP-122,
where the departure from linearity, that is associated with the destruction
of the Eu$^{2+}$ magnetic order at the surface is observed at similar
fluences (see Figs. \ref{fig:figAvsFEFA1eV55} and \ref{fig:figAvsFEFAP1eV55}).
The departure from linearity is followed by a complete saturation
above $\mathcal{F}\sim50$ $\mu$J/cm$^{2}$ where the temperature
within the complete probed volume exceeds the magnetic ordering $T$,
as expected from the saturation model\cite{KusarKabanov2008}, where
the saturation fluence value depends on the geometry and the optical
penetration depth of the beams. 

\begin{figure}[tbh]
\begin{centering}
\includegraphics[angle=-90,width=0.98\columnwidth]{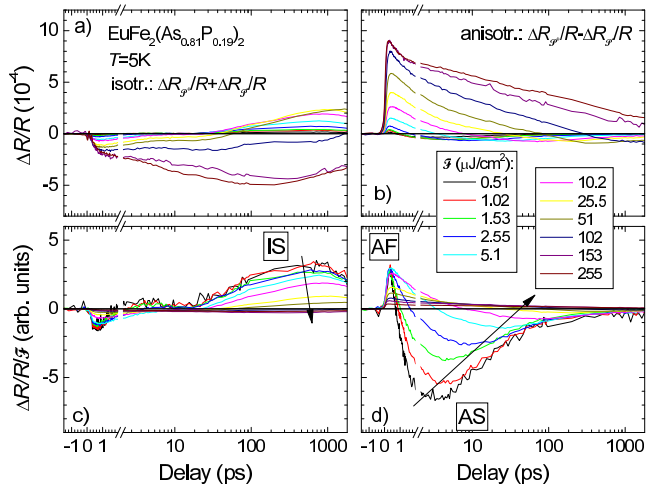} 
\par\end{centering}

\caption{(Color online) The isotropic, (a), and anisotropic, (b), transient
components in EuFe$_{2}$(As$_{0.81}$P$_{0.19}$)$_{2}$ at different
fluences obtained from the data shown in Fig. \ref{fig:figDRvsFEFAP1eV55}.
The $\mathcal{F}$-normalized transients from panels (a) and (b) are
shown in (c) and (d), respectively, to emphasize the low-$\mathcal{F}$
region. Arrows indicate the direction of increasing fluence. }

\label{fig:figDRvsFEFAP3eV1aniso} 
\end{figure}

In Eu-122 the saturation is, despite the same geometry of the beams,
observed above $\mathcal{F}\sim150$ $\mu$J/cm$^{2}$ only, which
is $\sim3$ times higher than in EuP-122. At this fluence the peak
surface lattice temperature is estimated to be $T_{\mathrm{\mathrm{TH}}}\sim70$K.
Deeper in the sample at a probe penetration depth $T_{\mathrm{TH}}\sim40$
K still significantly exceeds $T_{\mathrm{N}}$. The reason for the
difference could be associated with the slower risetime in Eu-122
that prevented us to measure the true amplitude of the response. Instead,
the $\Delta R/R$ value at the longest delay was measured, which is
still on the rising part of the signal, and depends on both, the amplitude
of the response and the rate of the magnetization suppression. It
is reasonable to expect that the rate depends on $T_{\mathrm{TH}}$
and as a result on the excitation density also for $T\mathrm{_{\mathrm{TH}}}>T_{\mathrm{N}}$.

Let us now focus on the complex fluence dependence of the transients
shape in the SC EuP-122. In the absence of an external magnetic field
at an intermediate $\mathcal{F}\sim$10 $\mu$J/cm$^{2}$ the slow
component appears rather isotropic (see Fig. \ref{fig:figDrVsT})
while the anisotropy was previously\cite{PogrebnaMertelj2015} observed
only in high magnetic field. The anisotropy of the slow component
observed at low excitation fluences in zero magnetic field (see Fig.
\ref{fig:figAvsFEFAP1eV55}) could therefore be a consequence of an
interplay with another anisotropic component that saturates at a rather
low $\mathcal{F}$ and is masked by other nonsaturated components
at increased $\mathcal{F}$. 

To test this hypothesis we calculate the isotropic part, $(\Delta R/R)_{\mathrm{iso}}=\nicefrac{1}{2}[(\Delta R/R)_{\mathcal{P^{^{+}}}}+(\Delta R/R)_{\mathcal{P^{^{-}}}}]$,
and the anisotropic part, $(\Delta R/R)_{\mathrm{an}}=\nicefrac{1}{2}[(\Delta R/R)_{\mathcal{P^{^{+}}}}-(\Delta R/R)_{\mathcal{P^{^{-}}}}]$,
of the data shown in Fig. \ref{fig:figDRvsFEFAP1eV55} and plot them
in Fig. \ref{fig:figDRvsFEFAP3eV1aniso}. 

The slow part of the obtained isotropic (IS) response is below $\mathcal{F\approx}100$
$\mu$J/cm$^{2}$ rather similar to the $\mathcal{P}^{+}$ response
in Eu-122 and $H$-parallel responses in high magnetic field\cite{PogrebnaMertelj2015}
in both samples and it can be associated with the out-of-plane Eu$^{2+}$
demagnetization upon the photoexcitation. 

\begin{figure}[tbh]
\begin{centering}
\includegraphics[angle=-90,width=0.98\columnwidth]{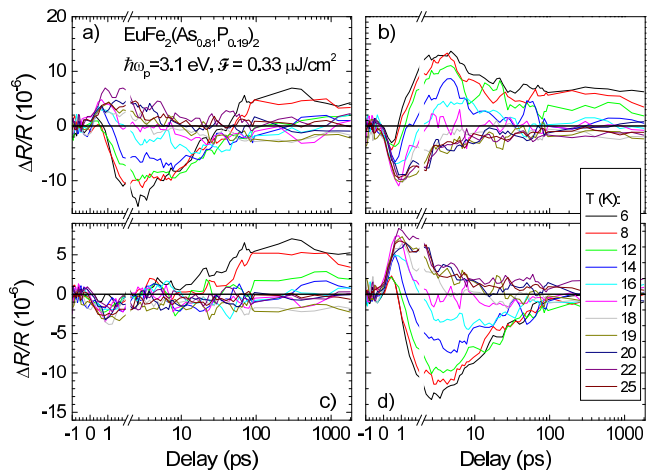} 
\par\end{centering}

\caption{(Color online) Photoinduced reflectivity transients at low $T$ and
extremely low $\mathcal{F}$ in EuP-122 for $\mathcal{P}^{+}$ (a)
and $\mathcal{P}^{-}$ (b) polarizations. The corresponding isotropic
and anisotropic transient components are shown in (c) and (d), respectively.
To reduce noise the traces were smoothed resulting in a reduction
of the time resolution to $\sim150$ fs. }

\label{fig:figEFAPvsTLowF3eV1} 
\end{figure}

\begin{figure}[tbh]
\begin{centering}
\includegraphics[angle=-90,width=0.7\columnwidth]{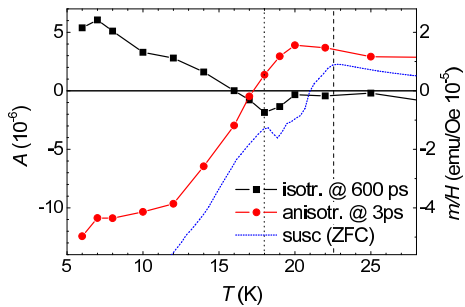} 
\par\end{centering}

\caption{(Color online) Temperature dependence of the amplitudes of the isotropic
part at 600-ps delay and the anisotropic part at 3-ps delay from Fig.
\ref{fig:figEFAPvsTLowF3eV1}, where the dotted and dashed vertical
lines indicate $T_{\mathrm{Curie}}$ and $T_{\mathrm{c}}$ (onset)
obtained form the susceptibility, respectively. The blue dotted line
is the ZFC susceptibility from Fig. \ref{fig:figSusc}.}

\label{fig:figAvsFEFAPlowF} 
\end{figure}

The anisotropic component of the low-$\mathcal{F}$ response is more
complex consisting from a fast positive component, marked AF in Fig.
\ref{fig:figDRvsFEFAP3eV1aniso} (d) and a slower negative component
marked AS, which is clearly observed only at low $\mathcal{F}$. 

While component AF can be clearly related to the Fe-$d$ SDW ordering
present below $T_{\mathrm{SDW}}\sim$35 K in our sample the origin
of component AS appears more elusive. In Ba(Fe,Co)$_{2}$As$_{2}$
a similar slowly relaxing anisotropic response was observed and clearly
associated with superconductivity.\cite{StojchevskaMertelj2012,TorchinskyMcIver2011}
In optimally doped Ba(Fe,Co)$_{2}$As$_{2}$ this SC response is completely
saturated above $\mathcal{F}_{\mathrm{sat}}\sim2$ $\mu$J/cm$^{2}$
due to the nonthermal destruction of the SC state,\cite{StojchevskaMertelj2012}
while in underdoped Ba(Fe,Co)$_{2}$As$_{2}$ and Sm(Fe,Co)AsO,\cite{MerteljStojchevska2013}
with lower $T_{\mathrm{c}}$s, $\mathcal{F}_{\mathrm{sat}}$s are
even lower and the magnitudes of the SC response smaller. 

Since component AS is similar to a possible anisotropic SC response
and the data in Fig. \ref{fig:figDRvsFEFAP3eV1aniso} (b) and (d)
suggest, that $\mathcal{F}_{\mathrm{sat}}^{\mathrm{AS}}$ is below
$0.5$ $\mu$J/cm$^{2}$, we measured $T$ dependence also at extremely
low $\mathcal{F}=0.33$ $\mu$J/cm$^{2}$ as shown in Fig. \ref{fig:figEFAPvsTLowF3eV1}.
Component AS vanishes with increasing $T$ at $\sim20$ K {[}see inset
to Fig. \ref{fig:figEFAPvsTLowF3eV1} (b){]}, significantly below
the bulk $T_{\mathrm{c}}$ (onset) of 22.7 K and rather close to $T_{\mathrm{N}}\sim18$
K. As a result, despite the similarity of component AS \foreignlanguage{british}{behaviour}
to the SC response in Ba(Fe,Co)$_{2}$As$_{2}$, a firm assignment
of component AS to the SC response is not possible. However, in Sm(Fe,Co)AsO\cite{MerteljStojchevska2013}
a similar discrepancy between the bulk $T_{\mathrm{c}}$ and the temperature
at which the transient SC component vanishes was observed so a tentative
assignment of the SC response to component AS in EuP-122 is plausible. 

Component AF, which corresponds to the Fe-$d$ SDW order, shows very
similar saturation \foreignlanguage{british}{behaviour} to component
IS with the linear $\mathcal{F}$-dependence up to a slightly higher
threshold fluence, $\mathcal{F}_{\mathrm{c}}^{\mathrm{SDW}}\sim20$
$\mu$J/cm$^{2}$. This fluence would correspond to a transient lattice
heating of $T_{\mathrm{TH}}\sim25$ K, that is just slightly lower
than the equilibrium SDW transition temperature {[}See Fig. \ref{fig:figAvsTfast}
(d){]}, $T_{\mathrm{SDW}}\sim35$ K, suggesting almost thermal destruction
of the SDW order. However, by taking into account: the fast sub-picosecond
$\mathcal{F}$-independent risetime, a limited accuracy of the transient-heating
estimate and the small lattice heat capacity in this $T$-range the
small difference between $T_{\mathrm{TH}}$ and $T_{\mathrm{SDW}}$
does not imply, that the thermal destruction of the SDW state is more
likely than a fast athermal subpicosecond melting of the SDW order. 

The relaxation time of component AF increases with increasing fluence.
At low fluences, where the SDW gap is not yet completely suppressed,
we attribute it to the bottleneck-governed SDW order recovery dynamics\cite{PogrebnaVujicic2014},
as in the undoped SDW iron pnictides\cite{StojchevskaKusar2010,StojchevskaMertelj2012,PogrebnaVujicic2014}.
At the fluences near and above the threshold the recovery slows down
to tens of picoseconds near the threshold to beyond a few hundred
picoseconds at the highest experimental fluence. On this timescales
the bottleneck mechanism can not be operative and the nematic lattice-strain
dynamics, that was observed recently\cite{PatzLi2014} to extend to
several 100 ps near the magnetostructural transition temperature,
might determine the characteristic time scale at these fluences. However,
on the nanosecond timescale the heat diffusion out of the experimental
volume also takes place. Since no clear evidence for a slow anisotropic
dynamics is observed in the low-excitation-density response at higher
$T$, where the nematic fluctuations dominate the response, the timescale
is most likely governed by heat diffusion.

\section{Summary and conclusions}

Transient optical reflectivity was investigated in EuFe$_{2}$(As$_{1-x}$P$_{x}$)$_{2}$
in the undoped, $x=0$, SDW and doped, $x=0.19$, SC state samples
as a function of the excitation fluence. The characteristic anisotropic
subpicosecond transient reflectivity response indicates the presence
of the SDW order below $T_{\mathrm{SDW}}\sim35$ K also in the $x=0.19$
SC sample suggesting \emph{a coexistence of Fe-$d$-SDW, SC and Eu$^{2+}-$FM}
\emph{orders}. At both dopings a characteristic bottleneck \foreignlanguage{british}{behaviour}
of the fast picosecond transient reflectivity response due to a partial
charge gap present in the SDW state was observed, consistent with
previous results.\cite{PogrebnaVujicic2014}

With increasing excitation pulse fluence the saturation of different
transient reflectivity components indicates a suppression of the SC
order below $\sim1$ $\mu$J/cm$^{2}$, followed by a suppression
of the\emph{ }Eu$^{2+}-$FM order at $\sim10$ $\mu$J/cm$^{2}$ (at
$T=1.5$ K). The SDW order is suppressed in the $x=0.19$ SC sample
at $\sim20$ $\mu$J/cm$^{2}$ while no suppression is evident in
the undoped $x=0$ sample up to $250$ $\mu$J/cm$^{2}$. 

While the suppression of the Eu$^{2+}-$FM order appears to be thermal
due to the lattice temperature rise above the respective magnetic
ordering temperatures, the SDW order suppression in the $x=0.19$
SC sample is nonthermal with the peak lattice temperature reaching
only $\sim10$ K below $T_{\mathrm{SDW}}$. The slow thermal suppression
of the Eu$^{2+}-$FM order further confirms the rather weak coupling\cite{PogrebnaMertelj2015}
of the Eu$^{2+}$ spins to the rest of the system.
\begin{acknowledgments}
Work at Jozef Stefan Institute was supported by ARRS (Grant No. P1-0040).
We would like to thank Z. Jaglicic for magnetic susceptibily measurements.
\end{acknowledgments}

\bibliography{biblio}

\end{document}